%% file: main.tex
\PassOptionsToPackage{table}{xcolor}
\documentclass[10pt,letterpaper]{article}

\usepackage[T1]{fontenc}
\usepackage{times}
\usepackage{helvet}
\usepackage{courier}
\usepackage[top=0.75in,bottom=0.85in,left=0.7in,right=0.7in,columnsep=0.28in]{geometry}
\usepackage[hyphens]{url}
\usepackage{graphicx}
\usepackage{natbib}
\usepackage{amsmath,amssymb}
\usepackage{xcolor}
\usepackage{colortbl}
\usepackage{listings}
\usepackage{float}
\usepackage{microtype}
\usepackage[hidelinks]{hyperref}
\hypersetup{
  pdftitle={Evaluating Predicted Densities, Hamiltonians, and Density Matrices as Periodic SCF Initializers},
  pdfauthor={Pin Chen, Jiang Li, Yutong Lu},
  pdfsubject={Preprint},
  pdfkeywords={density functional theory, SCF initialization, charge density, Hamiltonian, density matrix}
}

\newcommand{\abacus}{\textsc{ABACUS}}
\newcommand{\rhohd}{\ensuremath{\rho\mathrm{HD}\text{-}43\mathrm{K}}}

\title{Evaluating Predicted Densities, Hamiltonians, and Density Matrices\\as Periodic SCF Initializers}

\author{%
Pin Chen,\quad Jiang Li,\quad Yutong Lu\\
School of Computer Science and Engineering, Sun Yat-sen University, Guangzhou, China}

\date{}

\begin{document}

\twocolumn[{%
\maketitle
\vspace{-1.2em}
\begin{quote}
\textbf{Abstract.} Learned electronic states are usually evaluated by prediction error, even though their intended use is to accelerate the self-consistent-field (SCF) loop of density functional theory (DFT). We ask whether lower offline error actually yields a better SCF initializer. We construct \rhohd{}, a 43,851-crystal DFT corpus with aligned charge-density, Hamiltonian, and density-matrix labels, and extend the solver workflow to inject all three predicted states. The closed-loop benchmark compares direct converged-state prediction against residual prediction from solver-native references on a frozen test set of non-magnetic crystals, using paired convergence, iteration, and SCF-loop timing measurements. Injecting the exact converged density matrix or Hamiltonian as an oracle upper bound cuts the median SCF count from 16 to one, revealing substantial acceleration headroom. Among learned inputs, direct charge-density prediction (Charge3Net-E3) accelerates about 91\% of paired crystals, saves a median of three SCF iterations, and yields a \(1.18\times\) SCF-loop speedup. In contrast, the tested residual-density and matrix initializers do not consistently improve over the standard superposition-of-atomic-densities baseline. These results establish that target-space accuracy alone is insufficient: a learned initializer must also be compatible with the nonlinear solver trajectory, and its utility must be measured in the loop.
\end{quote}
\vspace{0.8em}
}]

\section{Introduction}

Density functional theory (DFT) underpins computational materials discovery, yet each calculation must solve a nonlinear self-consistent-field (SCF) problem. Modern neural models can predict charge densities, Hamiltonians, and density matrices directly from atomic structures. This suggests an appealing acceleration strategy: initialize SCF from a learned electronic state instead of the standard superposition-of-atomic-densities (SAD) guess. Most electronic-state models, however, are selected by target-space error---a criterion that does not measure the intended outcome. Small errors in solver-sensitive modes can alter mixing stability, occupations, or the basin reached by the iterative map, whereas even large errors in insensitive components may barely affect convergence.

Closed-loop evidence is emerging but remains fragmented across representations. In molecules, all three states have separately shortened SCF: learned Hamiltonians re-injected by QHFlow on QH9 \citep{yu2023qh9,kim2025qhflow}, learned densities on QM9 \citep{li2025superresolution}, and learned one-particle density matrices \citep{hazra2024density}. In crystals, by contrast, only density-based initialization has been demonstrated \citep{koker2024higher,chen2025ecd}. Yet no study compares all three states---charge density \(\rho\), Hamiltonian \(H\), and density matrix \(D\)---under a single periodic solver.

This gap is consequential. Periodic Hamiltonian models are usually assessed by matrix error, bands, or non-self-consistent properties, so whether a predicted \(H\) usefully initializes crystal SCF, how it differs from \(\rho\), and where \(D\) lies between them remain unknown. A fair test requires co-generating all three states with identical functional, pseudopotentials, basis, k points, and convergence settings, then re-injecting them into the same executable. This requirement motivates both \rhohd{} and our solver extensions.

We therefore study a direct question: \emph{which learned electronic states accelerate the DFT loop when used as initial conditions?} The relevant object is the SCF trajectory
\[
s_{\mathrm{SAD}} \rightarrow F(s_{\mathrm{SAD}}) \rightarrow \cdots \rightarrow s_*,
\]
where \(s\in\{\rho,D,H\}\). We treat the solver as part of the evaluation environment and compare all representations through the same downstream observables: convergence, SCF iterations, and SCF-loop time. This common currency separates two questions that offline benchmarks conflate: whether a state is close to the converged target, and whether it is a useful starting point for the numerical solver.

We evaluate two formulations. Direct models predict the converged state \(s_*\) from a structure \(x\). Residual models predict the correction from a solver-native reference: \(\rho_*-\rho_0\), \(D_*-D_0\), or \(H_*-H_0\), where \(\rho_0\) is the charge density after one SAD-started solver update. Residual learning is a plausible inductive bias because the reference already encodes atomic, basis, and pseudopotential information. Whether that bias improves SCF convergence is an empirical question, not an assumption of our benchmark.

\begin{figure}[t]
\centering
\includegraphics[width=\columnwidth]{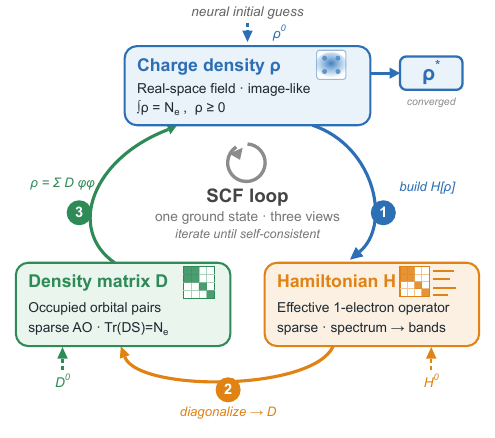}
\caption{One electronic ground state in three non-interchangeable representations: the image-like real-space charge density \(\rho\), and the orbital-based sparse matrices \(D\) and \(H\). The SCF loop cycles \(\rho\!\to\!H\!\to\!D\!\to\!\rho\) (Eq.~\ref{eq:state_cycle}), and a learned model can initialize any representation (\(\rho^0\), \(D^0\), \(H^0\)). The distinct information and constraints each carries are detailed in the Preliminaries.}
\label{fig:three_views}
\end{figure}

\begin{figure*}[t]
\centering
\includegraphics[width=0.98\textwidth]{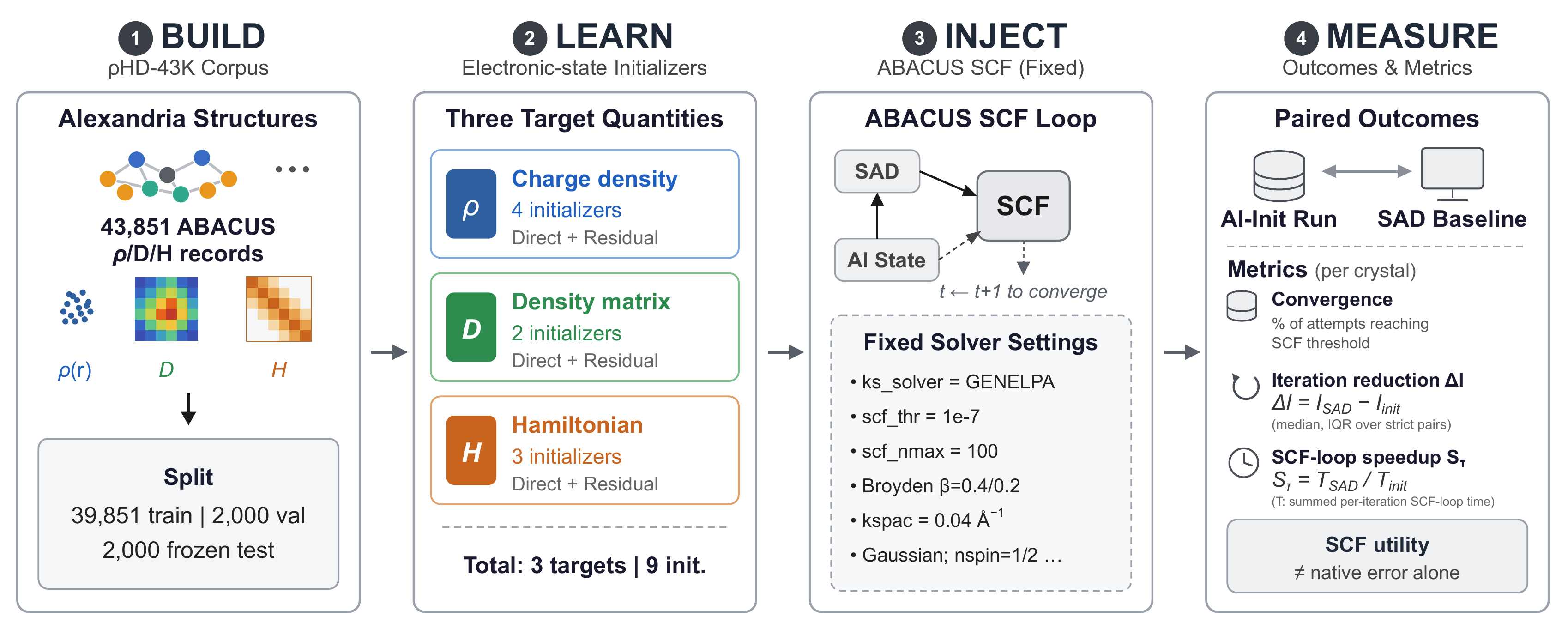}
\caption{Overview of the closed-loop benchmark. Alexandria structures are recomputed with \abacus{} to build \rhohd{}, an aligned \(\rho/H/D\) corpus. Nine learned initializers are injected through representation-specific \abacus{} interfaces. Every initialized run is paired with SAD for the same crystal under fixed solver settings, and converged-state oracle inputs quantify interface headroom. Prediction error measures state reconstruction; convergence, paired iteration reduction, and SCF-loop speedup determine whether a prediction accelerates DFT.}
\label{fig:overview}
\end{figure*}

Our benchmark uses periodic calculations with \abacus{}, an open-source DFT package \citep{zhou2025abacus}, and a localized numerical atomic-orbital basis. We construct \rhohd{} by selecting inorganic crystals from Alexandria \citep{schmidt2021alexandria} and recomputing aligned \(\rho\), \(H\), \(D\), and overlap labels under one electronic-structure protocol. A frozen test subset of about 2,000 structures supports the closed-loop study. It covers nine learned initializers spanning ELECTRAFI \citep{elsborg2025electra,elsborg2026global}, Charge3Net-E3 \citep{koker2024higher}, EdenGNN \citep{li2025differencecharge}, HamGNN \citep{zhong2023transferable}, DeepH-E3 \citep{gong2023deephe3}, and NextHAM \citep{yin2025advancing}. For each representation, we also re-inject the converged state as an oracle control. These controls distinguish a weak prediction from an initialization interface that cannot influence SCF. Figure~\ref{fig:overview} summarizes the benchmark pipeline.

Our contributions are:
\begin{itemize}
    \item \rhohd{}, a self-built corpus of 43,851 Alexandria-derived crystals recomputed under one \abacus{} protocol with aligned \(\rho\), \(H\), \(D\), and overlap data;
    \item AI-oriented \abacus{} initialization interfaces and a paired closed-loop protocol that evaluate nine learned initializers across all three representations through shared SCF outcomes, using converged-state oracle controls while retaining representation-specific native errors;
    \item evidence that model selection by native error alone is insufficient: direct charge prediction yields a modest but consistent speedup, whereas the tested residual and matrix models fail to realize their oracle headroom.
\end{itemize}

\section{Preliminaries: Three Representations of the Electronic State}

Kohn--Sham DFT obtains the ground-state density by solving a nonlinear fixed-point problem \citep{hohenberg1964inhomogeneous,kohn1965self}. In a non-orthogonal localized atomic-orbital (LCAO) basis, one SCF step links the three targets studied here (Fig.~\ref{fig:three_views}):
\begin{equation}
\begin{aligned}
H[\rho]C &= SC\varepsilon, \\
D &= CfC^\dagger, \\
\rho(\mathbf r) &= \sum_{\mu\nu\mathbf R}D_{\mu\nu}^{\mathbf R}
\phi_\mu(\mathbf r)\phi_\nu(\mathbf r-\mathbf R).
\end{aligned}
\label{eq:state_cycle}
\end{equation}
Here \(S\) is the overlap matrix, \(f\) contains occupations, and \(\mathbf R\) indexes periodic image blocks. The solver constructs \(H\) from \(\rho\), solves a generalized eigenproblem, forms \(D\), reconstructs a new \(\rho\), and mixes successive iterates until the input and output densities agree. Initialization and mixing therefore affect both stability and cost; Pulay and Broyden mixing methods accelerate the trajectory from its history \citep{pulay1980convergence,broyden1965class}, whereas learned initialization changes only its starting point.

The three targets carry the same electronic solution in different, non-interchangeable coordinates. A real-space density is a scalar field and is comparatively agnostic to the basis used to obtain it, although its grid grows with cell volume. It must integrate to the electron number and, in the physical convention used here, remain non-negative. The density matrix is compact and block sparse in an LCAO basis, but it is tied to element-specific orbitals and must satisfy Hermiticity, \(\mathrm{Tr}(DS)=N_e\), and approximate non-orthogonal idempotency \(DSD\approx D\) for an insulating zero-temperature state. The Hamiltonian is likewise a sparse covariant block matrix; its errors are amplified through the generalized eigensystem and occupations. Thus \(D\) and \(H\) are not merely ``small floating-point arrays.'' Elementwise errors can be small while violating a global subspace, trace, spectrum, or overlap-conditioned constraint that matters to SCF.

\section{Related Work}

\paragraph{Closed-loop evidence in molecules.}
QH9 contains Hamiltonians for 130,831 stable QM9 geometries and 2,399 molecular-dynamics trajectories, but evaluates prediction rather than SCF initialization \citep{yu2023qh9}. QHFlow later initialized PySCF from predicted \(H\) on 300 QH9 molecules, reporting 30\% fewer iterations and 13--15\% lower total runtime than QHNet after accounting for inference overhead \citep{kim2025qhflow}. A QM9-trained density ResNet saved, on average, 5.5 iterations relative to SAD initialization \citep{li2025superresolution}, and learned one-particle density matrices also shortened molecular SCF \citep{hazra2024density}. These separate studies establish molecule-level feasibility, not a controlled cross-state ranking.

\paragraph{Periodic density evidence.}
ELECTRA/ELECTRAFI \citep{elsborg2025electra,elsborg2026global} and ChargE3Net \citep{koker2024higher} learn transferable densities with equivariant representations. ChargE3Net reduced the median SCF count from 15 to 11 on unseen Materials Project crystals. ECD contains 140,646 PBE and 7,147 HSE structures; its VASP test on 19 common crystals achieved iteration ratios (initialized over default SCF iterations, lower is better) of \(0.757\) and \(0.681\) for the PBE-trained and HSE-tuned models \citep{chen2025ecd}. Periodic density acceleration is therefore established, although ECD's loop sample is small. EdenGNN instead reports difference-density reconstruction and non-self-consistent properties, without SCF reinjection \citep{li2025differencecharge}.

\paragraph{Periodic Hamiltonian evidence.}
DeepH predicts periodic DFT Hamiltonians \citep{li2022deeph}, and DeepH-E3 makes the mapping explicitly E(3)-equivariant \citep{gong2023deephe3}. HamGNN covers molecules and solids \citep{zhong2023transferable}; NextHAM learns \(H_*-H_0\) across materials \citep{yin2025advancing}. Their main evaluations use matrix MAE, bands, or direct inference that bypasses SCF, and their outputs remain tied to training-time orbitals. To our knowledge, learned periodic \(H\), \(D\), and \(\rho\) have not been compared as initializers on paired crystals.

To test whether the direct-Hamiltonian result is architecture-specific, we retrain DeepH-E3 on \rhohd{} with the same \abacus{} orbital basis and include it alongside HamGNN-Matrix-\(H\) and residual NextHAM. The adaptation retains the published equivariant architecture but adds an auditable \abacus{} data path, streaming inference, Hermitian block projection, and CSR export. This removes an implementation-only exclusion while preserving an important scope boundary: all three Hamiltonian models remain tied to the orbital convention used to generate their training labels.

\paragraph{Why a unified solver boundary is needed.}
PySCF accepts an initial AO density matrix \citep{sun2020pyscf}, SIESTA can reuse one \citep{garcia2020siesta}, and plane-wave codes commonly restart from densities or wavefunctions. Our claim is therefore comparative, not that matrix restart is unprecedented. Mixing codes, bases, pseudopotentials, or k points would confound state choice with protocol. We instead add checked \(D/H\) initialization to an \abacus{} 3.11.0-beta.4 LCAO workflow and place \(\rho\), \(D\), and \(H\) behind one periodic, MPI-capable boundary. To our knowledge, this is the first paired closed-loop comparison of all three learned states on inorganic crystals.

\section{Closed-Loop Formulation}

Let \(x\) be a crystal structure and \(F\) the SCF update implemented by the DFT solver. Starting from \(s_0\), the solver generates \(s_{t+1}=F(s_t;x)\) until it satisfies a fixed convergence threshold. We intervene only at initialization and leave the subsequent solver unchanged. The injected state is \(s\in\{\rho,D,H\}\).

\subsection{Direct and Residual Initialization}

A direct model predicts the converged state,
\begin{equation}
\hat{s}_{\mathrm{dir}}=f_\theta(x)\approx s_*.
\end{equation}
A residual model instead starts from a reference state \(s_{\mathrm{ref}}\) and predicts
\begin{equation}
\widehat{\Delta s}=g_\theta(x,s_{\mathrm{ref}}),\qquad
\hat{s}_{\mathrm{res}}=s_{\mathrm{ref}}+\widehat{\Delta s}.
\end{equation}
For \(\rho\), \(s_{\mathrm{ref}}=\rho_0=F_\rho(\rho_{\mathrm{SAD}};x)\), the charge after one SAD-started solver update. For \(D\), the reference is the one-step matrix \(D_0\) generated from the same SAD-started update. For NextHAM, it is the zeroth-step Hamiltonian \(H_0\). The comparison tests whether learning a nominally smaller correction produces a state that is more useful to SCF than predicting the converged state directly.

\subsection{Native and Closed-Loop Metrics}

\paragraph{Native accuracy.}
For charge density, we report normalized full-grid \(L_1\) error,
\begin{equation}
\epsilon_\rho =
\frac{\sum_{\mathbf{g}\in G} |\rho(\mathbf{g})-\hat{\rho}(\mathbf{g})|}
{\sum_{\mathbf{g}\in G} |\rho(\mathbf{g})|},
\end{equation}
where \(G\) is the real-space grid. For \(A\in\{D,H\}\), we report an elementwise MAE on a fixed, target-specific set of scalar LCAO entries \(\mathcal E_A\),
\begin{equation}
\epsilon_A =
\frac{1}{N_A}
\sum_{e\in\mathcal E_A}
\left| A_e-\hat{A}_e \right|,
\quad A\in\{D,H\},
\end{equation}
where \(N_A=|\mathcal E_A|\). For \(D\), \(\mathcal E_D\) contains the entries of the common symmetry-completed graph blocks. For \(H\), \(\mathcal E_H\) is the nonzero support of the converged Hamiltonian after all predictions have been symmetry-projected and reconstructed in the CSR form read by \abacus{}; a missing predicted entry is therefore evaluated as zero. The same support is used for all rows of a target. The scales and units of \(\epsilon_\rho\), \(\epsilon_D\), and \(\epsilon_H\) differ; only models for the same target can be compared by native error.

\paragraph{Closed-loop common currency.}
For each initializer, convergence rate is computed over all test-set attempts. Iteration and timing comparisons use strict pairs for which both the initialized and SAD runs converge. For a paired structure \(x\), we define
\begin{equation}
\begin{aligned}
\Delta I(x)&=I_{\mathrm{SAD}}(x)-I_{\mathrm{init}}(x),\\
S_T(x)&=\frac{T_{\mathrm{SAD}}(x)}{T_{\mathrm{init}}(x)}.
\end{aligned}
\end{equation}
Thus, \(\Delta I>0\) and \(S_T>1\) indicate acceleration. The strict-pair set \(\mathcal P\) contains only structures for which both runs converge and have valid iteration counts. We report convergence over all attempts and the median SCF iteration count of converged initialized runs. On strict pairs, we report median \(\Delta I\), its interquartile range (IQR), the number and percentage of faster pairs, and median \(S_T\). This combination exposes failures excluded by pairing while separating absolute iteration cost, paired improvement, and wall-time acceleration. \(T\) sums per-iteration SCF-loop times parsed from \abacus{} logs; it excludes neural inference, state export, process startup, and final-output I/O. Consequently, \(S_T\) measures solver-loop acceleration rather than end-to-end deployment speedup.

\section{Experimental Setup}

\subsection{Dataset and Solver}

\paragraph{\rhohd{} construction.}
\rhohd{} is our aligned electronic-state corpus, not a relabeling of an existing database. We select non-magnetic inorganic structures from Alexandria \citep{schmidt2021alexandria}, retain spin-unpolarized systems (\texttt{nspin=1}), and recompute every label with the same periodic \abacus{} LCAO workflow \citep{zhou2025abacus}.\footnote{The dataset, training and inference code, trained checkpoints, and the customized \abacus{} build that exports and re-reads learned states are provided as the Code and Data Supplement.} Each successful record couples the crystal and numerical settings to the converged real-space charge grid, Hamiltonian blocks, density-matrix blocks, and overlap matrix. Because all three target labels come from the same calculation, Eq.~\ref{eq:state_cycle} is numerically aligned rather than assembled across codes or basis sets.

\begin{figure*}[t]
\centering
\includegraphics[width=0.98\textwidth]{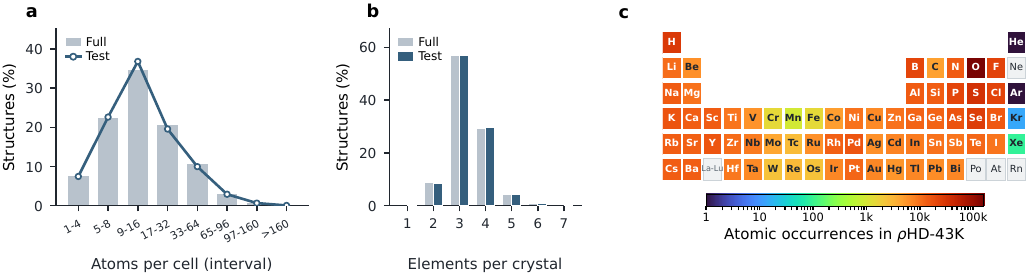}
\caption{Coverage of the full \rhohd{} corpus and its frozen closed-loop test set. (a) Distribution over explicit atom-count intervals per input cell; bars show the full corpus and the line shows the frozen test set. (b) Number of chemical species per crystal. (c) Element abundance in periodic-table layout; color encodes the total number of atomic occurrences in all \rhohd{} structures.}
\label{fig:hcd_coverage}
\end{figure*}

The successfully processed \rhohd{} pool contains 43,851 calculations, split into 39,851 training, 2,000 validation, and a frozen test set of 2,000 crystals (hereafter ``the test set''). One test crystal is excluded because its reference calculation did not converge within our compute budget, so all closed-loop results are reported over the remaining 1,999 evaluated crystals. Model selection uses only the training and validation partitions; all learned-model native errors and closed-loop claims use the same frozen test IDs. The SAD-derived matrix references cover all but two evaluated crystals, namely those with available \(\rho_0/D_0/H_0\) artifacts. Figure~\ref{fig:hcd_coverage} reports composition and cell-size coverage for both the full pool and the frozen test set, guarding against the concern that our acceleration results reflect only a chemically narrow subset.

The source calculations use the PBE exchange--correlation functional \citep{perdew1996generalized}, SG15 optimized norm-conserving pseudopotentials \citep{schlipf2015optimization}, and a double-zeta-plus-polarization numerical atomic-orbital basis. Closed-loop reruns preserve each structure's pseudopotentials, orbital files, k-point mesh, and parallel resources. All methods use \(10^{-7}\) as the SCF threshold, at most 100 iterations, Broyden mixing with \(\beta=0.2\), and one OpenMP thread per MPI process. The only changed input is the initialization mode and its associated state file. The complete solver input and reproducible workflow are provided in Appendix~\ref{app:code_release}.

\paragraph{AI-oriented \abacus{} interfaces.}
The benchmark uses an \abacus{} 3.11.0-beta.4-based build in which the LCAO initialization path accepts three state types under one input contract. \texttt{init\_chg file} reads a periodic charge grid; \texttt{init\_chg dm} reads real-space density-matrix blocks and reconstructs the initial charge; \texttt{init\_chg hr} reads Hamiltonian blocks, solves the initial generalized eigenproblem, and then resumes the standard SCF map. The matrix readers validate orbital dimensions and lattice-translation block layout before entering SCF. After initialization, diagonalization, occupation, mixing, convergence threshold, and output are identical to SAD. This design turns model evaluation into a controlled intervention rather than a separate surrogate calculation.

\subsection{Models and Initialization Interfaces}

The model set spans field and block-matrix representations while keeping the solver protocol fixed (Table~\ref{tab:model_selection}). For \(\rho\), ELECTRAFI and Charge3Net-E3 are direct predictors; a second Charge3Net-E3 model and EdenGNN predict difference densities. For localized matrices, HamGNN-Matrix predicts direct \(D_*\) and \(H_*\), and a separate head predicts \(\Delta D=D_*-D_0\). DeepH-E3 supplies an independent direct \(H_*\) predictor, while NextHAM predicts \(\Delta H=H_*-H_0\). Predicted fields are exported through \texttt{init\_chg file}; sparse matrices use the \texttt{init\_chg dm} and \texttt{init\_chg hr} interfaces.

The selected architectures are representative rather than exhaustive, with model families and citations given in the Introduction and Related Work. Each prediction is reconstructed into a complete solver input before either native or closed-loop evaluation, so residual rows never compare only the correction tensor. The design tests target and initialization choices under a common executable protocol; it is not an architecture leaderboard.

\begin{table}[t]
\centering
\caption{Learned initializers in the closed-loop benchmark.}
\label{tab:model_selection}
\scriptsize
\setlength{\tabcolsep}{3.2pt}
\begin{tabular}{llll}
\hline
Target & Model & Type & \texttt{init\_chg} \\
\hline
\(\rho\) & ELECTRAFI-\(\rho\) & direct & \texttt{file} \\
\(\rho\) & Charge3Net-E3-\(\rho\) & direct & \texttt{file} \\
\(\rho\) & Charge3Net-E3-\(\Delta\rho\) & residual & \texttt{file} \\
\(\rho\) & EdenGNN-\(\Delta\rho\) & residual & \texttt{file} \\
\hline
\(D\) & HamGNN-Matrix-\(D\) & direct & \texttt{dm} \\
\(D\) & HamGNN-Matrix-\(\Delta D\) & residual & \texttt{dm} \\
\hline
\(H\) & HamGNN-Matrix-\(H\) & direct & \texttt{hr} \\
\(H\) & DeepH-E3-\(H\) & direct & \texttt{hr} \\
\(H\) & NextHAM-\(\Delta H\) & residual & \texttt{hr} \\
\hline
\end{tabular}
\end{table}

\subsection{Baselines and Controls}

The default comparator is \texttt{init\_chg auto}, which constructs SAD without learned inference or export cost. For same-target native comparisons, \(\rho_0\) is the charge after one SAD-started update, \(H_0\) is the zeroth-step Hamiltonian assembled from SAD, and \(D_0\) is formed by its first diagonalization. The reference-generation cost is not included in the reported residual-model SCF-loop timing. These are solver-derived references: element, pseudopotential, basis, and electrostatic information has already been supplied analytically before the learned correction is applied.

We additionally re-inject each converged test target through the same file used by its learned counterpart. These oracle runs jointly validate serialization and quantify interface headroom; they are diagnostic upper bounds, not deployable baselines. Every initialized run is paired with the SAD run for the same structure. Table~\ref{tab:main} reports convergence over all attempts and iteration/time statistics only over strict converged pairs. The native SAD matrix artifacts are available for all but two test crystals, explaining the reference-set exception noted in Table~\ref{tab:native_accuracy}.

\section{Results}

\begin{figure*}[t]
\centering
\includegraphics[width=0.98\textwidth]{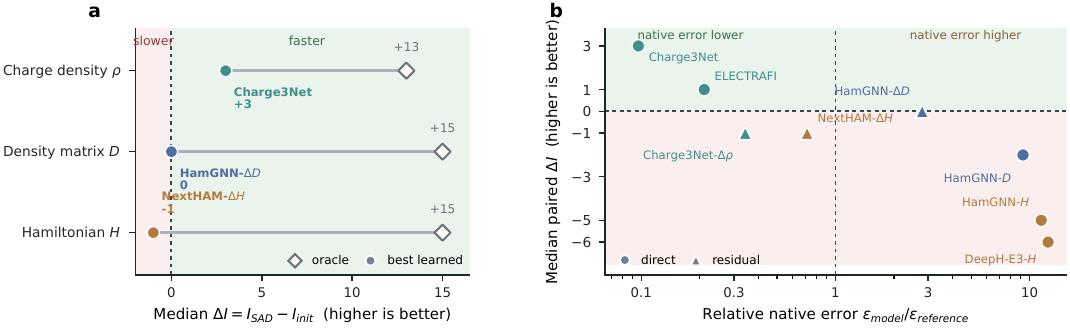}
\caption{Offline accuracy versus SCF utility. (a) Oracle and best-learned median \(\Delta I\) for each interface; positive values are faster than SAD. (b) Native error relative to the corresponding SAD-derived reference versus median paired \(\Delta I\), including both direct-\(H\) architectures, HamGNN-Matrix-\(H\) and DeepH-E3-\(H\). Colors denote \(\rho\), \(D\), \(H\); circles and triangles denote direct and residual formulations. EdenGNN-\(\Delta\rho\) overlaps Charge3Net-E3-\(\Delta\rho\) and is omitted for clarity.}
\label{fig:closed_loop_summary}
\end{figure*}

We separate prediction quality from solver utility. Table~\ref{tab:native_accuracy} evaluates each reconstructed state against \(s_*\), Table~\ref{tab:main} evaluates the corresponding model family inside SCF, and Figure~\ref{fig:closed_loop_summary}(b) relates the two. Residual rows always refer to the reconstructed initializer, not the residual tensor alone.

\subsection{Native Accuracy Is Target-Specific}

\begin{table*}[t]
\centering
\caption{Native error against the converged state \(s_*\) on the frozen test set (lower is better). Errors are comparable within, but not across, targets. Density rows use 1,999 crystals; all matrix rows use the common 1,997-crystal subset with \(D_0/H_0\). Hamiltonian rows use the converged-state support after CSR reconstruction.}
\label{tab:native_accuracy}
\footnotesize
\renewcommand{\arraystretch}{1.10}
\begin{tabular}{llllr}
\hline
Target & Model / reference & Initializer form & Native metric & Test error \\
\hline
\rowcolor{gray!12}
\(\rho\) & SAD-derived reference & \(\rho_0\) & full-grid \(\epsilon_\rho\) & \(4.508\times10^{-2}\) \\
\(\rho\) & ELECTRAFI-\(\rho\) & \(\hat{\rho}_*\) & full-grid \(\epsilon_\rho\) & \(9.510\times10^{-3}\) \\
\(\rho\) & Charge3Net-E3-\(\rho\) & \(\hat{\rho}_*\) & full-grid \(\epsilon_\rho\) & \(4.354\times10^{-3}\) \\
\(\rho\) & Charge3Net-E3-\(\Delta\rho\) & \(\rho_0+\widehat{\Delta\rho}\) & full-grid \(\epsilon_\rho\) & \(1.546\times10^{-2}\) \\
\(\rho\) & EdenGNN-\(\Delta\rho\) & \(\rho_0+\widehat{\Delta\rho}\) & full-grid \(\epsilon_\rho\) & \(1.671\times10^{-2}\) \\
\hline
\rowcolor{gray!12}
\(D\) & SAD-derived reference & \(D_0\) & block MAE & \(5.832\times10^{-5}\) \\
\(D\) & HamGNN-Matrix-\(D\) & \(\hat{D}_*\) & block MAE & \(5.389\times10^{-4}\) \\
\(D\) & HamGNN-Matrix-\(\Delta D\) & \(D_0+\widehat{\Delta D}\) & block MAE & \(1.629\times10^{-4}\) \\
\hline
\rowcolor{gray!12}
\(H\) & SAD-derived reference & \(H_0\) & support MAE & \(2.133\times10^{-4}\) \\
\(H\) & HamGNN-Matrix-\(H\) & \(\hat{H}_*\) & support MAE & \(2.446\times10^{-3}\) \\
\(H\) & DeepH-E3-\(H\) & \(\hat{H}_*\) & support MAE & \(2.652\times10^{-3}\) \\
\(H\) & NextHAM-\(\Delta H\) & \(H_0+\widehat{\Delta H}\) & support MAE & \(1.524\times10^{-4}\) \\
\hline
\end{tabular}
\end{table*}

As reported in Table~\ref{tab:native_accuracy}, all four learned charge densities improve over the SAD-derived one-step reference \(\rho_0\) in native error, by factors from \(2.7\times\) (EdenGNN) to \(10.4\times\) (direct Charge3Net-E3). The matrix ordering is more nuanced. Direct \(D\) is \(9.2\times\) worse than \(D_0\), while residual \(D\) narrows the gap to \(2.8\times\). Direct HamGNN-\(H\) and DeepH-E3-\(H\) are respectively \(11.5\times\) and \(12.4\times\) worse than \(H_0\). In contrast, reconstructed NextHAM-\(\Delta H\) reaches \(0.71\times\) the \(H_0\) error, improving the solver-native reference offline. Its closed-loop result below nevertheless has no median acceleration, providing a particularly direct counterexample to selection by native MAE alone.

\subsection{Oracle Headroom and Learned Acceleration}

\begin{table*}[t]
\centering
\caption{Closed-loop SCF performance on the frozen test set (one formal result per crystal and initializer; pre-execution infrastructure retries are excluded). Oracle rows inject the exact converged state as an upper-bound control, not a deployable method. Conv. is over all results; Med. \(I\), Med. \(\Delta I\) (IQR), faster-pair counts, and SCF-loop speedup use strict pairs.}
\label{tab:main}
\footnotesize
\renewcommand{\arraystretch}{1.10}
\setlength{\tabcolsep}{3.4pt}
\begin{tabular}{lllrrrrr}
\hline
Target & Model / initializer & Type & Conv. & Med. \(I\) & Med. \(\Delta I\) [IQR] & Faster, \(n\) (\%) & SCF spd. \\
\hline
\rowcolor{gray!12}
\(\rho,D,H\) & \abacus{} SAD default & baseline & 99.70\% & 16 & \(0\ [0,0]\) & -- & \(1.00\times\) \\
\(\rho\) & True charge density & oracle & 99.70\% & 3 & \(+13\ [+12,+15]\) & 1993 (100.0) & \(5.30\times\) \\
\(\rho\) & ELECTRAFI-\(\rho\) & direct & 99.25\% & 15 & \(+1\ [0,+2]\) & 1162 (58.6) & \(1.06\times\) \\
\(\rho\) & Charge3Net-E3-\(\rho\) & direct & 99.40\% & 14 & \(+3\ [+2,+3]\) & 1806 (90.9) & \(1.18\times\) \\
\(\rho\) & Charge3Net-E3-\(\Delta\rho\) & residual & 99.20\% & 18 & \(-1\ [-3,0]\) & 367 (18.5) & \(0.93\times\) \\
\(\rho\) & EdenGNN-\(\Delta\rho\) & residual & 99.30\% & 18 & \(-1\ [-3,0]\) & 367 (18.5) & \(0.93\times\) \\
\hline
\(D\) & True density matrix & oracle & 99.70\% & 1 & \(+15\ [+13,+17]\) & 1993 (100.0) & \(14.56\times\) \\
\(D\) & HamGNN-Matrix-\(D\) & direct & 99.70\% & 18 & \(-2\ [-3,-1]\) & 90 (4.5) & \(0.90\times\) \\
\(D\) & HamGNN-Matrix-\(\Delta D\) & residual & 99.65\% & 16 & \(0\ [-1,+1]\) & 845 (42.4) & \(1.00\times\) \\
\hline
\(H\) & True Hamiltonian & oracle & 99.70\% & 1 & \(+15\ [+13,+17]\) & 1993 (100.0) & \(15.19\times\) \\
\(H\) & HamGNN-Matrix-\(H\) & direct & 99.60\% & 22 & \(-5\ [-8,-3]\) & 0 (0.0) & \(0.75\times\) \\
\(H\) & DeepH-E3-\(H\) & direct & 99.65\% & 23 & \(-6\ [-10,-4]\) & 1 (0.05) & \(0.73\times\) \\
\(H\) & NextHAM-\(\Delta H\) & residual & 99.55\% & 17 & \(-1\ [-3,+1]\) & 710 (35.7) & \(0.97\times\) \\
\hline
\end{tabular}
\end{table*}

Table~\ref{tab:main} reports one formal result per test crystal for every initializer. Missing final SCF logs, runtime crashes after executable launch, and runs reaching \texttt{scf\_nmax=100} are counted as failures and remain in the convergence denominator. Scheduler or interconnect failures before \abacus{} starts may be retried and are recorded separately. The 28 existing final directories without SCF logs comprise 26 segmentation faults and two MPI/GLEX failures. The six oracle failures are shared with SAD and reach 100 iterations, so the appropriate oracle result is 1,993 converged calculations plus six closed-loop failures rather than an artificially forced perfect score.

The oracle controls show that all three interfaces can materially change convergence. Exact \(\rho\) reduces the median from 16 to three iterations and gives a \(5.30\times\) loop speedup. Exact \(D\) and \(H\) reach one-iteration medians and speedups of \(14.56\times\) and \(15.19\times\), respectively. All 1,993 converged oracle pairs are faster than SAD, validating the three initialization interfaces while retaining the six shared failures.

Among learned initializers, direct Charge3Net-E3 gives the clearest gain: 1,806 of 1,987 pairs (90.9\%) use fewer iterations, with median \(\Delta I=+3\), IQR \([+2,+3]\), and \(1.18\times\) loop speedup. ELECTRAFI gives a smaller improvement (median \(\Delta I=+1\), \(1.06\times\)). In contrast, both residual-density models have median \(\Delta I=-1\) and speedups below one. Figure~\ref{fig:closed_loop_summary}(a) places the best learned initializer for each interface against its oracle ceiling.

Learned matrices do not realize their much larger oracle ceiling. Direct \(D\) matches SAD's 99.70\% convergence but has median \(\Delta I=-2\), with 1,639 of 1,993 strict pairs slower. The two direct \(H\) architectures agree qualitatively: HamGNN-\(H\) accelerates no strict pair and has median \(\Delta I=-5\), while DeepH-E3-\(H\) accelerates one of 1,992 pairs (0.05\%), has median \(\Delta I=-6\), and yields a \(0.73\times\) loop speedup. Six of its seven non-converged cases are shared with SAD; one is model-only. Residual \(D\) is effectively neutral (median \(\Delta I=0\), \(1.00\times\)), whereas NextHAM has median \(\Delta I=-1\) and \(0.97\times\) speedup despite beating \(H_0\) in native error.

\subsection{Offline Error Does Not Determine SCF Outcome}

The native-error orderings of Table~\ref{tab:native_accuracy} do not carry over to the loop. Figure~\ref{fig:closed_loop_summary}(b) normalizes each model's native error by its corresponding solver-native reference and plots it against median paired \(\Delta I\). The direct density models lie on the accelerating side of \(\Delta I=0\), whereas both residual-density models lie on the slowing side despite improving over \(\rho_0\) offline. Direct \(D/H\) models remain above their same-target references and yield no median gain; NextHAM crosses below the offline reference but still slows the median loop. EdenGNN-\(\Delta\rho\) is omitted from the panel because it overlaps Charge3Net-E3-\(\Delta\rho\); both remain reported in Table~\ref{tab:main}. Across the nine initializers, native error neither ranks objectives at the model level nor predicts individual SCF outcomes.

\section{Discussion}

\paragraph{The baselines are not equally informed.}
The field reference \(\rho_0\) starts from the simple SAD superposition plus one solver update, whereas \(H_0\) is assembled directly from the structure, pseudopotentials, basis, and SAD potential, and \(D_0\) from solving its generalized eigenproblem. These material-specific references expose different amounts of solver structure to the learned correction, so a learned density competes with a comparatively crude field reference while a learned matrix must compete with a highly structured orbital prior. This explains why both direct \(H\) architectures are much worse than \(H_0\). NextHAM is the informative exception: its residual correction beats \(H_0\) in support MAE but still fails to accelerate the median loop, showing that a stronger prior explains the offline ordering without making offline proximity sufficient.

\paragraph{Matrices carry structure that elementwise error obscures.}
Beyond the cross-representation scale mismatch noted in Results, matrix initialization imposes global constraints that elementwise MAE cannot capture: \(D\) must respect the non-orthogonal overlap, electron count, occupied subspace, and approximate idempotency, and \(H\) must yield a consistent spectrum and density after diagonalization. Because the oracle \(D\) and \(H\) interfaces are highly effective when these relationships are correct, the closed-loop shortfall reflects solver-compatible prediction rather than lack of interface leverage.

\paragraph{Residual learning is not automatically solver-aligned.}
A smaller target magnitude does not guarantee an easier downstream problem: residual fields can contain chemically specific, high-frequency corrections, and an inaccurate correction can move the SAD-derived \(\rho_0\) state away from the solver basin. This explains why the residual models improve on \(\rho_0\) in density error yet still slow SCF.

\section{Limitations}

Our conclusions are scoped to one in-distribution, non-magnetic Alexandria corpus and one \abacus{} LCAO workflow; the ranking may shift with a different basis, functional, mixing scheme, magnetism, or metallic smearing, and we do not test out-of-distribution transfer to surfaces, defects, or strongly correlated systems. Finally, our reported speedups measure SCF-loop time and exclude model inference, state export, and residual-reference generation. For the direct charge-density winner the excluded cost is essentially inference (a few minutes per structure), small against the far longer wall-clock time of a periodic \abacus{} SCF run---commonly at the hour scale---so the \(1.18\times\) loop speedup should largely persist end to end. A fully audited end-to-end benchmark is left to future work.

\section{Conclusion}

We reframed learned electronic states as SCF initial conditions rather than regression targets, and built \rhohd{} with interfaces that inject predicted \(\rho\), \(D\), and \(H\) into one periodic \abacus{} workflow, pairing every run with SAD under fixed settings so each model becomes a controlled intervention read directly from the loop.

The closed-loop view overturns the offline ranking: injecting the exact converged \(D\) or \(H\) collapses the median SCF count to a single iteration, yet among learned inputs only direct charge-density prediction accelerates SCF consistently, while residual-density and learned-matrix initializers do not reliably beat SAD despite comparable or lower native error. Offline accuracy is thus neither necessary nor sufficient for SCF utility: models should be trained and reported by their SCF effect, and claims should rest on paired convergence, iteration, and loop-timing statistics. We release \rhohd{} and the closed-loop interfaces to make solver-in-the-loop evaluation standard; closing the \(D\)/\(H\) oracle gap and showing out-of-distribution acceleration are natural next steps.

\small
\bibliographystyle{plainnat}
\bibliography{references}

\clearpage
\appendix
\setcounter{secnumdepth}{2}
\setcounter{table}{0}
\renewcommand{\thetable}{A\arabic{table}}
\setcounter{figure}{0}
\renewcommand{\thefigure}{A\arabic{figure}}
\setcounter{equation}{0}
\renewcommand{\theequation}{A\arabic{equation}}
\makeatletter
\setlength{\@dblfptop}{0pt}
\makeatother
\renewcommand{\theHequation}{A.\arabic{equation}}
\renewcommand{\theHtable}{A.\arabic{table}}
\renewcommand{\theHfigure}{A.\arabic{figure}}
\input{appendix}

\end{document}

%% file: appendix.tex
\section{Reproducibility and Implementation Details}
\label{app:reproducibility}

This appendix records the dataset checks, model hyperparameters, selected training states, implementation changes, native-error audits, and solver-export rules used for the reported benchmark. Parameter counts denote trainable scalar parameters and exclude optimizer states and non-trainable buffers.

\subsection{Solver Configuration and Reproducible Workflow}
\label{app:code_release}

Every SAD, learned, and oracle run shares one \abacus{} LCAO input contract; only the initialization mode and its state file change between runs. Listing~\ref{lst:abacus_input} gives the complete \texttt{INPUT} used for the closed-loop reruns (directory paths are anonymized). All artifacts---the ABACUSflow workflow driver that reproduces label generation and the closed-loop reruns, the customized \abacus{} build that exports and re-reads learned \(\rho\), \(D\), and \(H\) states, and the \rhohd{} dataset together with the model training code and inference scripts---are provided as the Code and Data Supplement.

\begin{lstlisting}[
  float=h,
  basicstyle=\ttfamily\scriptsize,
  columns=fixed,
  keepspaces=true,
  frame=single,
  aboveskip=6pt,belowskip=6pt,
  captionpos=b,
  label={lst:abacus_input},
  caption={Complete \abacus{} \texttt{INPUT} for the closed-loop calculations. Pseudopotential and orbital directories are anonymized; only the \texttt{init\_chg}/state-file settings vary across initializers.}
]
INPUT_PARAMETERS
suffix       Scf-Spinoff-H   out_chg          1
pseudo_dir   <pp_dir>        ks_solver        genelpa
orbital_dir  <orb_dir>       basis_type       lcao
calculation  scf             out_hsr_npz      1
ntype        3               out_dm_npz       1
nspin        1               out_mul          1
symmetry     0               smearing_method  gaussian
kspacing     0.04            smearing_sigma   0.001
ecutwfc      120             mixing_type      broyden
scf_thr      1e-07           mixing_beta      0.2
scf_nmax     100             gamma_only       0
\end{lstlisting}

\subsection{Dataset Audit and Preprocessing}

The successful \rhohd{} pool contains 43,851 spin-unpolarized calculations, split into 39,851 training, 2,000 validation, and 2,000 test structures. Split membership is fixed before model training. One test crystal is excluded because its reference calculation did not converge within the compute budget, leaving 1,999 evaluated crystals. The density, Hamiltonian, density-matrix, and overlap labels for an individual structure come from the same converged \abacus{} calculation.

For matrix learning, each periodic atom-pair block \(M_{ij}^{\mathbf R}\), \(M\in\{H,D,S\}\), is accompanied by its inverse block
\begin{equation}
M_{ji}^{-\mathbf R}=\left(M_{ij}^{\mathbf R}\right)^{\mathsf T}.
\end{equation}
Missing inverse entries in the raw sparse graph are symmetry-completed before training, and fixed graph caches are accepted only when the inverse-missing count is zero. The same element-specific orbital definition and a maximum of 40 numerical atomic orbitals per atom are used when converting labels and predictions. Two evaluated crystals lack the complete one-step matrix artifacts required for \(D_0/H_0\), so matrix-reference native errors use 1,997 structures; closed-loop convergence uses all 1,999 attempted directories.

\subsection{Architectures and Parameter Counts}

Table~\ref{tab:appendix_architecture} gives the instantiated architectures rather than only the model-family names. Direct and residual versions of Charge3Net-E3 share the same network size, as do the three HamGNN-Matrix predictors; their objectives and reconstruction rules differ. DeepH-E3 is trained independently and contributes a second direct-Hamiltonian architecture.

\begin{table*}[t]
\centering
\caption{Architecture settings and trainable parameter counts for the nine learned initializers. Cutoffs are in \AA.}
\label{tab:appendix_architecture}
\scriptsize
\renewcommand{\arraystretch}{1.13}
\setlength{\tabcolsep}{3.4pt}
\begin{tabular}{llllp{9.4cm}}
\hline
Target & Model & Form & Parameters & Main instantiated settings \\
\hline
\(\rho\) & ELECTRAFI & direct & 40.166M & EScAIP backbone; 2 interaction layers, width 256, 32 attention heads; 30-\AA{} cutoff, at most 250 neighbors, 120 Gaussian density primitives per electron. \\
\(\rho\) & Charge3Net-E3 & direct & 1.909M & E(3)-equivariant density model; 3 interactions, multiplicity 500, \(l_{\max}=4\), 4-\AA{} cutoff, 20 neighbors, and 20 Gaussian radial functions. \\
\(\rho\) & Charge3Net-E3 & residual & 1.909M & Same architecture as the direct model; predicts \(\Delta\rho=\rho_*-\rho_0\). \\
\(\rho\) & EdenGNN & residual & 0.825M & 3 interactions, 4-\AA{} cutoff, 20 neighbors, \(l_{\max}=6\), 8 radial functions, element embedding 8, 4 density channels, and radial hidden width 64. \\
\(D\) & HamGNN-Matrix & direct & 2.903M & 3 equivariant layers; 26-\AA{} cutoff; 64 radial functions; node irreps \(64\!\times\!0e+64\!\times\!0o+32\!\times\!1o+16\!\times\!1e+8\!\times\!2o+16\!\times\!2e+8\!\times\!3o+8\!\times\!3e\). \\
\(D\) & HamGNN-Matrix & residual & 2.903M & Same backbone; predicts \(\Delta D=D_*-D_0\) with the one-step matrix supplied as a sidecar. \\
\(H\) & HamGNN-Matrix & direct & 2.903M & Same backbone and block decoder as the density-matrix models; direct converged-Hamiltonian target. \\
\(H\) & DeepH-E3 & direct & 4.544M & E(3)-equivariant Hamiltonian network; 3 interaction blocks, 26-\AA{} cutoff, spherical harmonics through \(l=4\), 64 scalar embedding channels, and one global vocabulary with element-specific orbital definitions for all 67 \rhohd{} elements. \\
\(H\) & NextHAM & residual & 0.842M & Conditional graph-attention transformer; 4 layers, 4 attention heads, hidden/equivariant features through \(l=6\), spherical harmonics through \(l=5\), 8-\AA{} radius, 64 radial functions, and zero dropout. \\
\hline
\end{tabular}
\end{table*}

EdenGNN uses the following hidden representation:
\begin{multline}
64\!\times\!0e+64\!\times\!0o+20\!\times\!1o+20\!\times\!1e+16\!\times\!2o+16\!\times\!2e\\
{}+8\!\times\!3o+8\!\times\!3e+6\!\times\!4o+6\!\times\!4e+5\!\times\!5e+5\!\times\!5o+4\!\times\!6e+4\!\times\!6o.
\end{multline}

\subsection{Optimization and Selected Training States}

Table~\ref{tab:appendix_training} lists the effective settings of the runs used for closed-loop inference. Batch size is per GPU. ``Probes'' denotes randomly sampled real-space query points per structure for density training; matrix models use one structure per GPU step.

\begin{table*}[t]
\centering
\caption{Training settings and model states used for closed-loop inference. The selected step is read from checkpoint metadata rather than inferred from a directory name.}
\label{tab:appendix_training}
\scriptsize
\renewcommand{\arraystretch}{1.13}
\setlength{\tabcolsep}{3.0pt}
\begin{tabular}{llrcllp{4.1cm}}
\hline
Model & Optimizer & Initial LR & GPUs & Batch / probes & Closed-loop state & Objective and regularization \\
\hline
ELECTRAFI-\(\rho\) & Muon+AdamW & \(10^{-4}\) & 8 & 1 / -- & epoch 28 & Electron-normalized integrated \(L_1\); final restart segment used \(10^{-5}\) LR and gradient clipping at 0.1. \\
Charge3Net-E3-\(\rho\) & Adam & \(10^{-3}\) & 4 & 6 / 500 & step 99,935 & Pointwise \(L_1\); power-decay schedule \((\alpha=0.96,\ \beta=3000)\). \\
Charge3Net-E3-\(\Delta\rho\) & Adam & \(5\!\times\!10^{-4}\) & 4 & 4 / 2,048 & step 200,000 & Electron-normalized residual-density \(L_1\); same power-decay schedule. \\
EdenGNN-\(\Delta\rho\) & AdamW & \(10^{-3}\) & 4 & 1 / 2,048 & step 297,900 & Solver-aligned residual loss: density weight 1 and neutrality weight 0.1; weight decay \(10^{-6}\). \\
HamGNN-Matrix-\(D\) & AdamW & \(10^{-4}\) & 2 & 1 / -- & step 405,000 & Matrix MAE plus Hermiticity 0.1, idempotency 0.01, trace 50, onsite 1, and onsite-diagonal 10. \\
HamGNN-Matrix-\(\Delta D\) & AdamW & \(10^{-4}\) & 4 & 1 / -- & step 443,000 & Residual MAE plus Hermiticity 0.1, idempotency 0.01, trace 0.1, onsite 1, and onsite-diagonal 10. \\
HamGNN-Matrix-\(H\) & AdamW & \(10^{-4}\) & 4 & 1 / -- & step 365,000 & Matrix MAE with onsite weight 2, onsite-diagonal weight 10, and large-entry weight 3 capped at 10. \\
DeepH-E3-\(H\) & Adam & \(10^{-4}\) & 4 & 1 / -- & epoch 27 & Equivariant direct-Hamiltonian block objective over all supported orbital-block types; lowest completed validation loss at the inference snapshot. \\
NextHAM-\(\Delta H\) & Adam & \(10^{-4}\) & 4 & 1 / -- & epoch 70; step 688,810 & Residual-Hamiltonian MAE; 39,357 size-filtered training structures and at most 100 validation structures per validation pass. \\
\hline
\end{tabular}
\end{table*}

Charge3Net-E3 uses 1,000 validation probes per structure for the direct run and 8,192 for the residual run. EdenGNN uses 4,096 validation probes. HamGNN-Matrix uses a plateau scheduler with gradient clipping at 1.0; direct \(H\) allows up to 300 epochs, while the two \(D\) runs allow 70. DeepH-E3 uses Adam betas \((0.9,0.999)\), a revert-and-decay scheduler with patience 15 and factor 0.8, and seed 42. EdenGNN uses a plateau scheduler with factor 0.5, patience 5, and minimum learning rate \(10^{-6}\). All Charge3Net-E3 and EdenGNN runs use seed 42; NextHAM uses seed 7.

\subsection{Changes Relative to the Released Model Implementations}

We separate changes to the scientific model from changes required to place its output behind the common solver boundary. Our modified code for every model is released with the benchmark (Appendix~\ref{app:code_release}); below we also link each upstream implementation we adapted, where an official one is available.

\paragraph{ELECTRAFI.}
The EScAIP density architecture is retained. We add a \rhohd{} valence-electron adapter, periodic neighbor capping for large cells, robust tiling when the number of Gaussian modes exceeds the feature width, finite-value checks, synchronized invalid-batch skipping under distributed training, and deterministic checkpoint export. These are dataset, scale, and training-stability changes rather than a new density architecture.

\paragraph{Charge3Net-E3.}
The direct model retains the released three-interaction E(3)-equivariant architecture.\footnote{Upstream implementation: \url{https://github.com/AIforGreatGood/charge3net}.} The residual variant keeps that architecture but changes the target to \(\rho_*-\rho_0\), increases the number of sampled probes, and replaces pointwise \(L_1\) by an electron-normalized residual loss. Both variants use an \abacus{} restart-data loader and exporter.

\paragraph{EdenGNN.}
This is a substantive implementation adaptation. No official code is released, so we reimplement the method from its original publication \citep{li2025differencecharge}. We implement the paper-described difference-density branch in the local periodic density interface, restrict it to the pseudo-density channel appropriate to the present norm-conserving LCAO calculations, and add a neutrality penalty. No PAW augmentation density is used. The trained output is reconstructed with the one-step reference before solver injection.

\paragraph{HamGNN-Matrix.}
The equivariant HamGNN backbone\footnote{Upstream implementation: \url{https://github.com/QuantumLab-ZY/HamGNN}.} is shared across \(H\), \(D\), and \(\Delta D\). The main changes are an \abacus{} orbital-block definition, symmetry-completed periodic graphs, a generalized real-space CSR writer, matrix-specific losses, and physical penalties for \(D\). The residual \(D\) model additionally reads \(D_0\) as a sidecar and exports \(D_0+\widehat{\Delta D}\). Thus the backbone is not redesigned, but the target, constraints, and solver-facing output are materially different from a standard Hamiltonian-only run.

\paragraph{DeepH-E3.}
The released implementation\footnote{Upstream implementation: \url{https://github.com/Xiaoxun-Gong/DeepH-E3}.} is configurable for a selected chemistry, but its original preprocessing assumes a fixed element vocabulary and atomic-orbital layout: the element map is inferred from the first graph and the orbital table from the first structure. This assumption is appropriate for element-specific datasets, but it cannot safely represent \rhohd{}, in which different crystals contain different subsets and combinations drawn from 67 elements. In particular, an element absent from the first structure can otherwise receive an invalid index, and Hamiltonian labels with composition-dependent orbital dimensions cannot be collated consistently.

We retain the E(3)-equivariant message-passing and Hamiltonian-decoding core, while extending its chemistry and data interfaces as follows. First, all fixed-split structures are scanned to construct a deterministic, dataset-wide map from physical atomic numbers to model indices. Second, the per-atom \abacus{} orbital metadata are reduced to one canonical orbital definition per element; preprocessing stops if the same atomic number is associated with inconsistent numerical orbitals. Third, each structure's Hamiltonian tensor is padded to the largest orbital dimension in the global basis, while a structure-specific mask restricts the loss to element pairs and orbital blocks that are actually present. Fourth, the original ordered element-pair one-hot vector, whose width scales as \(S^2\) for \(S\) species, is replaced by the concatenated identities of the two edge endpoints, whose width is \(2S\). This preserves edge direction and removes the quadratic chemistry bottleneck without changing the equivariant interaction mechanism. Fifth, the global atomic-number map and orbital table are stored with the checkpoint. Evaluation structures are remapped into that vocabulary, and inference fails explicitly for an unseen element or an incompatible orbital definition rather than silently constructing a malformed Hamiltonian.

To make full-corpus training practical, we additionally use the frozen train/validation/test lists, lazy per-structure graph and Hamiltonian loading, distributed data-parallel training, and masks that skip target blocks absent from a minibatch. Here ``general-element'' therefore means that a single checkpoint can be trained on heterogeneous crystals containing arbitrary subsets of the 67 supported \rhohd{} elements. It does not claim zero-shot transfer to an element, pseudopotential, or numerical-orbital basis absent from training. The solver-facing adaptation then streams one structure at a time during inference, supplies \abacus{} structures and orbital metadata, averages forward/reverse periodic blocks, projects the reconstructed matrix to Hermiticity, and exports real-space CSR blocks. These implementation changes are included in the released third-party patch.

\paragraph{NextHAM.}
This is the largest model-side adaptation. The released graph-attention network\footnote{Upstream implementation: \url{https://github.com/DavidYin94/NextHAM}.} is used as a conditional residual corrector: \(H_0\) is mapped into a global 66-element orbital basis, passed through the weak-Hamiltonian input branch, and the network predicts \(H_*-H_0\). Inference reconstructs \(H_0+\widehat{\Delta H}\), projects inverse periodic block pairs, enforces Hermiticity, and writes the \abacus{} CSR representation. Consequently, this row should be interpreted as a NextHAM-derived conditional \abacus{} initializer rather than an unchanged off-the-shelf checkpoint.

\subsection{Reconstruction and Solver Export}

All native errors for residual models are evaluated after reconstructing the full initializer:
\begin{equation}
\hat\rho=\rho_0+\widehat{\Delta\rho},\qquad
\hat D=D_0+\widehat{\Delta D},\qquad
\hat H=H_0+\widehat{\Delta H}.
\end{equation}
For the Charge3Net-E3 and EdenGNN density paths, the reconstructed real-space field is clipped at zero and rescaled to the reference electron count before export. The binary charge restart stores the real-space grid in double precision and its reciprocal representation in complex double precision. Matrix outputs are symmetry-completed, checked against the structure-specific orbital dimensions and translation-block layout, and serialized as real-space CSR blocks. The \(H\) interface solves the initial generalized eigenproblem after reading the predicted blocks; the \(D\) interface reconstructs the initial charge from the read density matrix.

The one-step references \(\rho_0\) and \(D_0\), and the zeroth-step \(H_0\), are precomputed from the solver-native SAD path. Their generation, neural inference, conversion, and file I/O are excluded from \(T\). Therefore Table~\ref{tab:main} is deliberately a solver-loop comparison. In particular, the residual-density methods are already slower than SAD even before charging the extra one-step reference cost; they cannot yield a positive end-to-end speedup under the present implementation.

\subsection{Closed-Loop Failure and Pairing Rules}

Each initializer has 1,999 formal material results. A run is counted as failed if its final SCF log is absent, the executable crashes after launch, or the solver reaches \texttt{scf\_nmax=100}. Such cases remain in the convergence-rate denominator. Pre-execution scheduler or interconnect failures may be retried and are logged separately. DeepH-E3 had two GLEX endpoint failures before \abacus{} started; both retries completed, so neither is counted as an SCF failure. Iteration differences, faster-pair counts, and loop-time speedups use only strict pairs for which both the initialized and SAD runs converge and expose valid iteration/timing records. This pairing prevents changes in the evaluated material subset from masquerading as acceleration.

For archival reproducibility, the released benchmark package (Appendix~\ref{app:code_release}) provides the frozen split IDs, malformed-file blacklist, graph-cache schema and inverse-pair audit, the trained checkpoints and their hashes, per-structure native errors, per-run convergence records, and the exact \abacus{} executable hash.

\subsection{DeepH-E3 Native and Closed-Loop Audit}

Native error is measured at two boundaries because the model tensor and the actual solver input are not identical. On DeepH-E3's 6,129,854 predicted HDF5 blocks (2,179,620,603 scalar entries), the checkpoint-matched block-weighted MAE against the converged labels is \(4.409\times10^{-3}\). These blocks cover 60.37\% of the blocks stored in the target HDF5 files; blocks outside the prediction graph are absent. After forward/reverse averaging, Hermitian projection, thresholding, and complete CSR reconstruction, we evaluate all Hamiltonian initializers on the same nonzero support and common 1,997-crystal subset available for \(H_0\). DeepH-E3 then has support MAE \(2.652\times10^{-3}\). The corresponding values are \(2.446\times10^{-3}\) for HamGNN-Matrix-\(H\), \(1.524\times10^{-4}\) for reconstructed NextHAM-\(\Delta H\), and \(2.133\times10^{-4}\) for \(H_0\). This common exported-input metric is the value reported in Table~\ref{tab:native_accuracy}.

For audit completeness, the zero-padded dense CSR MAEs on this subset are \(4.681\times10^{-4}\), \(4.317\times10^{-4}\), and \(2.690\times10^{-5}\) for DeepH-E3, HamGNN, and NextHAM, respectively. Dense MAE is smaller because the denominator includes many zero entries and is therefore not used for model selection or the main-paper comparison. The distinction also explains why the raw HDF5 block MAE, the reconstructed support MAE, and the dense MAE must not be placed in one column without qualification.

\begin{table}[t]
\centering
\caption{Checkpoint-matched DeepH-E3-\(H\) closed-loop audit. Achieved ratio is \(I_{\mathrm{init}}/I_{\mathrm{SAD}}\); speedup is \(T_{\mathrm{SAD}}/T_{\mathrm{init}}\).}
\label{tab:appendix_deephe3_scf}
\footnotesize
\renewcommand{\arraystretch}{1.08}
\setlength{\tabcolsep}{4pt}
\begin{tabular}{lr}
\hline
Metric & Value \\
\hline
Converged & 1,992 / 1,999 (99.65\%) \\
Strict converged pairs & 1,992 \\
Median initialized iterations & 23 \\
Median \(\Delta I\) [IQR] & \(-6\ [-10,-4]\) \\
Faster / same / slower pairs & 1 / 5 / 1,986 \\
Faster-pair fraction & 0.05\% \\
Median / aggregate achieved ratio & 1.375 / 1.475 \\
Median / aggregate SCF-loop speedup & 0.727 / 0.650 \\
\hline
\end{tabular}
\end{table}

Six of the seven non-converged DeepH-E3 structures also fail from SAD; only \texttt{agm003251875} is model-only. Over the 1,992 strict pairs, DeepH-E3 uses 49,224 SCF iterations versus 33,373 for SAD. Its per-structure support MAE has Spearman correlation 0.188 with achieved ratio (0.251 for dense MAE): statistically associated with worse behavior, but too weak to serve as a standalone acceleration criterion.

\subsection{Checkpoint Provenance and Audit Artifacts}

Checkpoint metadata, not directory labels, gives step 99,935 for direct Charge3Net-E3, step 200,000 for residual Charge3Net-E3, step 297,900 for EdenGNN, and epoch 28 for ELECTRAFI. All of these checkpoints are provided in the released package (Appendix~\ref{app:code_release}).

The direct-Hamiltonian closed loops use the symmetry-fixed, focus-loss HamGNN-Matrix checkpoint at step 365,000 and the DeepH-E3 checkpoint at epoch 27. The latter contains 4,543,772 trainable scalars, has snapshot validation loss \(2.111\times10^{-4}\), and has SHA256 prefix \texttt{cb2fc10c70a1afb0}. The frozen 1,999-ID list has SHA256 prefix \texttt{b2c5793a977232d9}. Full hashes are stored in the released \texttt{SHA256SUMS} and run metadata. The native and closed-loop numbers reported here are checkpoint-matched. Per-structure HDF5 and CSR errors, convergence records, exceptional IDs, and aggregation summaries are included in the released audit artifacts.